\documentclass[a4paper,11pt]{article}
\usepackage{pos}
\usepackage{comment}
\usepackage{lineno}
\usepackage{amsmath}
\usepackage{mathtools}
\usepackage{csquotes}
\usepackage{subcaption}
\usepackage[belowskip=-15pt,aboveskip=0pt]{caption}

\let\OLDthebibliography\thebibliography
\renewcommand\thebibliography[1]{
  \OLDthebibliography{#1}
  \setlength{\parskip}{0pt}
  \setlength{\itemsep}{0pt plus 0.3ex}
}

\newcommand\ddfrac[2]{\frac{\displaystyle #1}{\displaystyle #2}}

\title{Muon deficit in simulations of air showers inferred from AGASA data}

\ShortTitle{Muon deficit from AGASA data}

\author*[a,b]{Flavia Gesualdi}
\author[a]{Daniel Supanitsky}
\author[a]{Alberto Etchegoyen}
\affiliation[a]{Instituto de Tecnologías en Detección y Astropartículas (CNEA, CONICET, UNSAM), San Martín, Buenos Aires, Argentina}
\affiliation[b]{Karlsruhe Institute of Technology, Institute for Astroparticle Physics, Karlsruhe, Germany}
\emailAdd{flavia.gesualdi@iteda.cnea.gov.ar}

\abstract{Multiple experiments reported evidences of a muon deficit in air shower simulations with respect to data, which increases with the primary energy. In this work, we study the muon deficit using measurements of the muon density at $1000\,$m from the shower axis obtained by the Akeno Giant Air Shower Array (AGASA). The selected events have reconstructed energies in the range $18.83\,\leq\,\log_{10}(E_{R}/\textrm{eV})\,\leq\,19.46$ and zenith angles $\theta\leq 36^{\circ}$. We compare these muon density measurements to proton, iron, and mixed composition scenarios, obtained by using the high-energy hadronic interaction models EPOS-LHC, QGSJetII-04, and Sibyll2.3c. 
We find that AGASA data are compatible with a heavier composition, lying above the predictions of the mixed composition scenarios. The average muon density divided by the energy in AGASA data is greater than in the mixed composition scenarios by a factor of $1.49\pm0.11\,\textrm{(stat)}\pm^{0.49}_{0.30}\,\textrm{(syst)}$, $1.54\pm0.12\,\textrm{(stat)}\pm^{0.50}_{0.31}\,\textrm{(syst)}$, and $1.66\pm0.13\,\textrm{(stat)} \pm ^{0.54}_{0.34}\,\textrm{(syst)}$ for EPOS-LHC, Sibyll2.3c, and QGSJetII-04, respectively. We interpret this as further evidence of a muon deficit in air shower simulations at the highest energies.}

\FullConference{37$^{\rm{th}}$ International Cosmic Ray Conference (ICRC 2021)\\
		July 12th -- 23rd, 2021\\
		Online -- Berlin, Germany}

\begin{document}
\maketitle
\section{Introduction}
It is well known that high-energy cosmic rays can only be studied through the extensive air showers they produce, because the flux drops steeply above $10^{15}\,$eV. These primaries can reach energies beyond those accessible at the Large Hadron Collider. Therefore, high-energy hadronic interaction models can only be tested at ultrahigh energies via air showers. Furthermore, to elucidate the origin of cosmic rays, it is important to know the mass composition as a function of the energy. The composition is determined through the comparison of data to air shower simulations. The two most sensitive observables to the composition are the depth of the shower maximum $X_{\text{max}}$ and the number of muons $N_{\mu}$ \cite{Prado19}.

The muonic component of air showers, which originates mainly from the decay of hadrons, serves as a tracer of the hadronic component because most muons (and anti-muons) reach the ground before decaying. Recently, several experiments reported a muon deficit in simulations with respect to data. At the energies considered in this work, the Pierre Auger Observatory reported a deficit between $30\,\%$ and $80\,\%$ against the mixed composition scenarios \cite{Auger15,Auger16}. The Telescope Array Collaboration reported a deficit of $\sim 67\,\%$ against proton QGSJetII-04 showers (compatible with their observed composition) \cite{TelescopeArray18}. Finally, preliminary data of Yakutsk Array are compatible with no muon deficit \cite{Dembinski19, Cazon19}. It is not yet understood the cause of said deficit: it could be due to new high-energy phenomena, or to a partial mismodeling of hadronic interactions at low or high energies. Solving the muon deficit would improve both hadronic interaction models and the systematic uncertainties in the inferred comsic-ray mass composition.

In this work, we use data measured by the Akeno Giant Air Shower Array (AGASA) to study the muon deficit. This experiment was able to detect air showers with energies above $\sim 3\times10^{18}\,$eV and zenith angles $\theta\leq 45^{\circ}$. It consisted of 111 scintillation counters, covering $\sim100\,\text{m}^{2}$, and 27 proportional counters shielded with $30\,$cm of iron or $1\,$m of concrete, covering $\sim30\,\text{km}^{2}$. The latter served as the muon detectors, and had a vertical muon energy threshold of $0.5\,$GeV \cite{Hayashida95}. The experiment was decomissioned in 2004.

\section{Analysis}
We analyze the AGASA measurements of the muon density at $1000\,$m from the shower axis, measured on the shower plane. The data is extracted from Fig. 7 of Ref. \cite{Shinozaki04} (see also Table IV in Appendix B of Ref. \cite{Gesualdi20}).  The energies of these events are reduced by a factor of $0.68$ to take them to the referece energy scale defined by the \textit{Spectrum working group} \cite{Ivanov17}. This factor is obtained by matching the features of the cosmic ray flux measurements (see Ref.~\cite{Gesualdi20} and references therein). The idea behind matching fluxes is that, assuming isotropy, all experiments should measure the same flux. Furthermore, most of the difference between the AGASA and the reference energy scale arises from the muon deficit in the simulations used to calibrate the AGASA scale (especially in those using older-generation hadronic models). Considering the energies in the reference scale, the analyzed events have reconstructed energies in $18.83 \leq \log_{10} (E_{R}/\text{eV}) \leq 19.46$, as well as zenith angles $\theta \leq 36^{\circ}$.

Moreover, we use a library of proton, helium, nitrogen, and iron-initiated air showers, simulated using EPOS-LHC, QGSJetII-04, and Sibyll2.3c, as well as Fluka 2011.2x for the low-energy interactions, and employing CORSIKA. The library consists of $\sim20$ showers ($\sim30$ in the case of proton primaries) per model, primary, and input energy, the latter taking discrete values in the range $18.0 \leq \log_{10}(E/\text{eV}) \leq 19.8$ in steps of $\Delta \log_{10} (E/\text{eV}) = 0.2$. This library is further described in Ref.~\cite{Gesualdi20}. For every primary and model, the mean muon density as a function of the energy $\langle \widetilde{\rho}_{\mu}\rangle$ is fitted using a power-law. 

We derive a simulated muon density for a mixed composition scenario by using the mass fractions obtained from the fits to the $X_{\text{max}}$ distributions of the Pierre Auger Observatory \cite{Bellido17}. For each model, we compute the mixed composition mean muon density as $\langle \widetilde{\rho}_{\mu, \text{mix}} \rangle(E)= \sum_{A} f_{A}(E) \langle \widetilde{\rho}_{\mu,A} \rangle(E)$, where $A = \{ \text{p, He, N, Fe} \}$, $f_{A}(E)$ are the mass fractions, and $\langle\widetilde{\rho}_{\mu,A} \rangle(E)$ are the single-primary mean muon densities.

The mean muon density from simulations at $1000\,$m from the shower axis $\langle \widetilde{\rho}_{\mu}\rangle$, which depends on the Monte-Carlo energy $E$, cannot be directly compared to that of data, which depends on the reconstructed energy $E_R$, even if $E_R$ is an unbiased estimator of $E$ \cite{Dembinski18}. We therefore compute the energy reconstruction and binning effects on simulations analytically. The mean simulated muon density divided by the reconstructed energy as a function of the center of the $i$-th reconstructed energy bin $E_{Ri}$ can be expressed as

\begin{equation}
\left\langle\ddfrac{\rho_{\mu}}{E_R}\right\rangle(E_{Ri}) = \ddfrac{\int_{E_{Ri}^{-}}^{E_{Ri}^{+}} dE_{R}\, \int_{0}^{\infty} dE\,
\langle \widetilde{\rho}_{\mu}\rangle(E) \, E_R^{-1} \, J(E)\, G(E_{R}|E) }{\int_{E_{Ri}^{-}}^{E_{Ri}^{+}} dE_{R}\, 
\int_{0}^{\infty} dE \, J(E)\, G(E_{R}|E)},
\label{eq:conv}
\end{equation}
%
where $E_{Ri}^{-}$ and $E_{Ri}^{+}$ are the lower and upper limits of the energy bin; $\langle \widetilde{\rho}_\mu(E)\rangle$ is the mean muon density as a function of the Monte-Carlo/true energy of the simulation (i.e. the power-law fits); $J(E)$ is the cosmic ray flux, obtained by fitting the Telescope Array measurements \cite{TAFlux}; and $G (E_{R}|E)$ is the conditional probability distribution of $E_R$ conditioned to $E$, modeled as a log-normal distribution \cite{Takeda03} with a standard deviation that decreases with energy \cite{Yoshida95}. A more extensive explanation of Eq.~(\ref{eq:conv}) can be found in Ref.~\cite{Gesualdi20}.

Both the energy reconstruction and binning have the effect of making $\left\langle \rho_\mu/E_{R} \right\rangle(E_{Ri}=E^{*})$ smaller than $\langle \widetilde{\rho}_\mu/E\rangle(E = E^{*})$, when evaluating both at a same numerical value $E^{*}$. The energy reconstruction effect, which is the dominant, increases for broader conditioned distributions $ G(E_{R}|E)$ and in regions where the flux $J(E)$ is steeper. 

\begin{figure}[h]
\begin{center}
\includegraphics[width=0.75\textwidth]{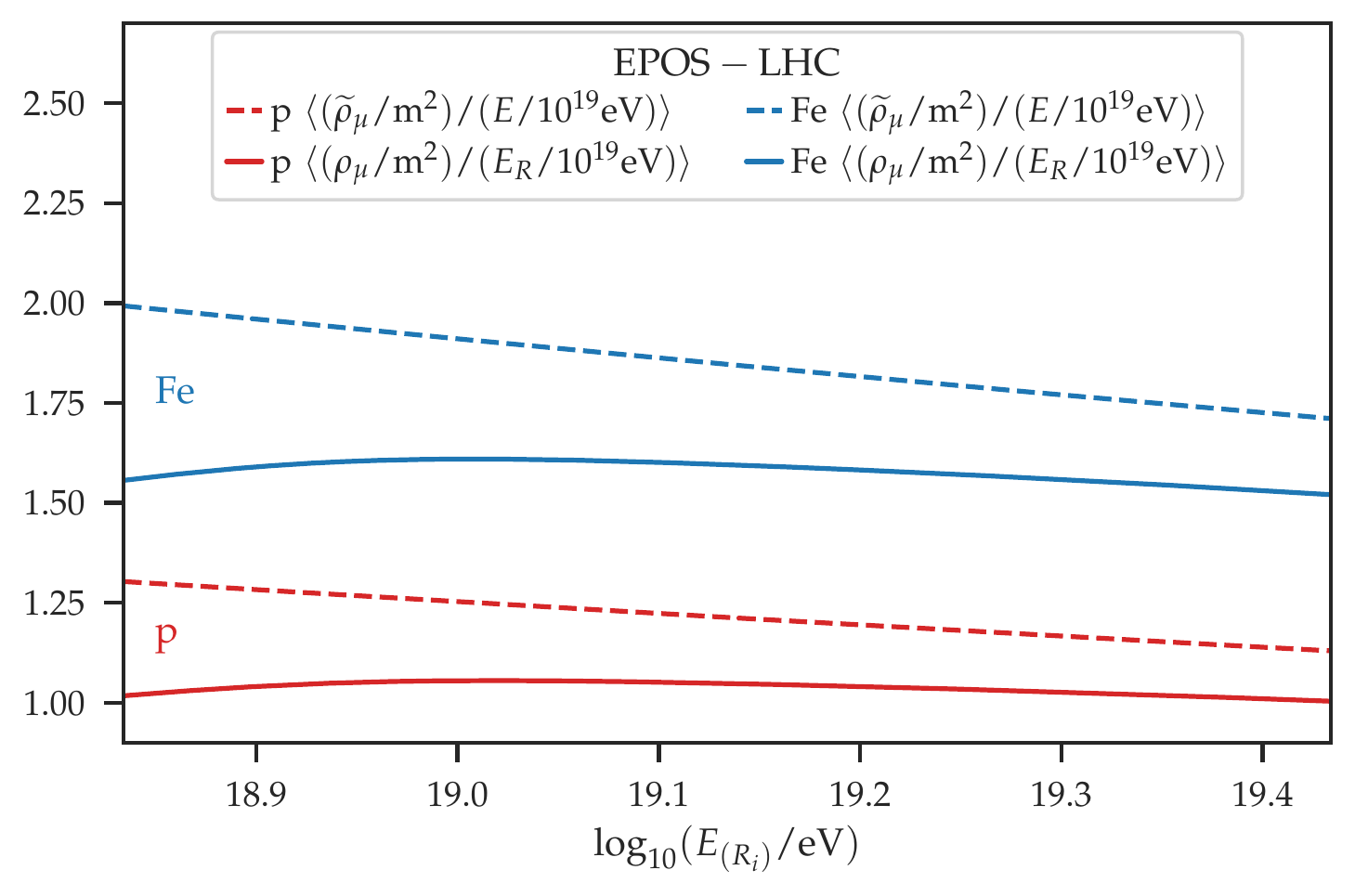}  
\caption{Mean muon density divided by the energy (reconstructed energy), as a function of the logarithm of the Monte-Carlo energy (center of the $i$-th reconstructed energy bin) in dashed lines (solid lines), obtained from the power-law fits (from Eq.~(\ref{eq:conv})). The shown muon densities correspond to proton and iron-initiated simulated air showers generated by using EPOS-LHC. The binwidth considered in Eq.~(\ref{eq:conv}) is $\Delta \log_{10} (E_{R}/\text{eV}) = 0.2$.
\label{fig:conv}}
\end{center}
\end{figure}

Figure \ref{fig:conv} shows a comparison between $\langle\widetilde{\rho}_{\mu}/E\rangle(E)$ and $\langle\rho_{\mu}/E_R\rangle(E_{Ri})$. It can be seen in the figure that $\langle\rho_{\mu}/E_R\rangle(E_{Ri})$ can be $11\,\%$ to $22\,\%$ smaller than 
$\langle \widetilde{\rho}_{\mu}/E\rangle(E)$ in the studied energy range. At low energies, this difference is explained by the large uncertainty 
in the reconstructed energy ($\sim\! 28\,\%$ at $10^{18.83}\,\textrm{eV}$). At high energies, the dominant effect 
is the flux suppression. The effect of binning in reconstructed energy with a bin width of $\Delta\log_{10}(E_R/\textrm{eV})=0.2$ is small in comparison.

It is relevant to add that $\left\langle\rho_{\mu}/E_R\right\rangle(E_{Ri})$ is computed from the integration of $\langle \widetilde{\rho}_{\mu, \text{mix}} \rangle(E)$ using Eq.~(\ref{eq:conv}). It is not equivalent to calculate it as the linear combination of $\left\langle\rho_{\mu,A}/E_R\right\rangle(E_{Ri})$, because the mass fractions $f_{A}(E)$ depend on the energy $E$, which is an integration variable in Eq.~(\ref{eq:conv}).

Finally, we consider the systematic uncertainties as presented in Ref.~\cite{Gesualdi21}. The systematic uncertainty in the energy constitutes the dominant source, ranging between $\pm 22\,\%$ to $\pm 26\,\%$ (this includes a $\pm 10\,\%$ systematic uncertainty of the reference energy scale). The systematics in the muon density are negligible in comparison. The propagation of uncertainties is applied following Ref.~\cite{Gesualdi20}.

\section{Results}

\begin{center}
\begin{figure}[!ht]
\includegraphics[width=0.5\linewidth]{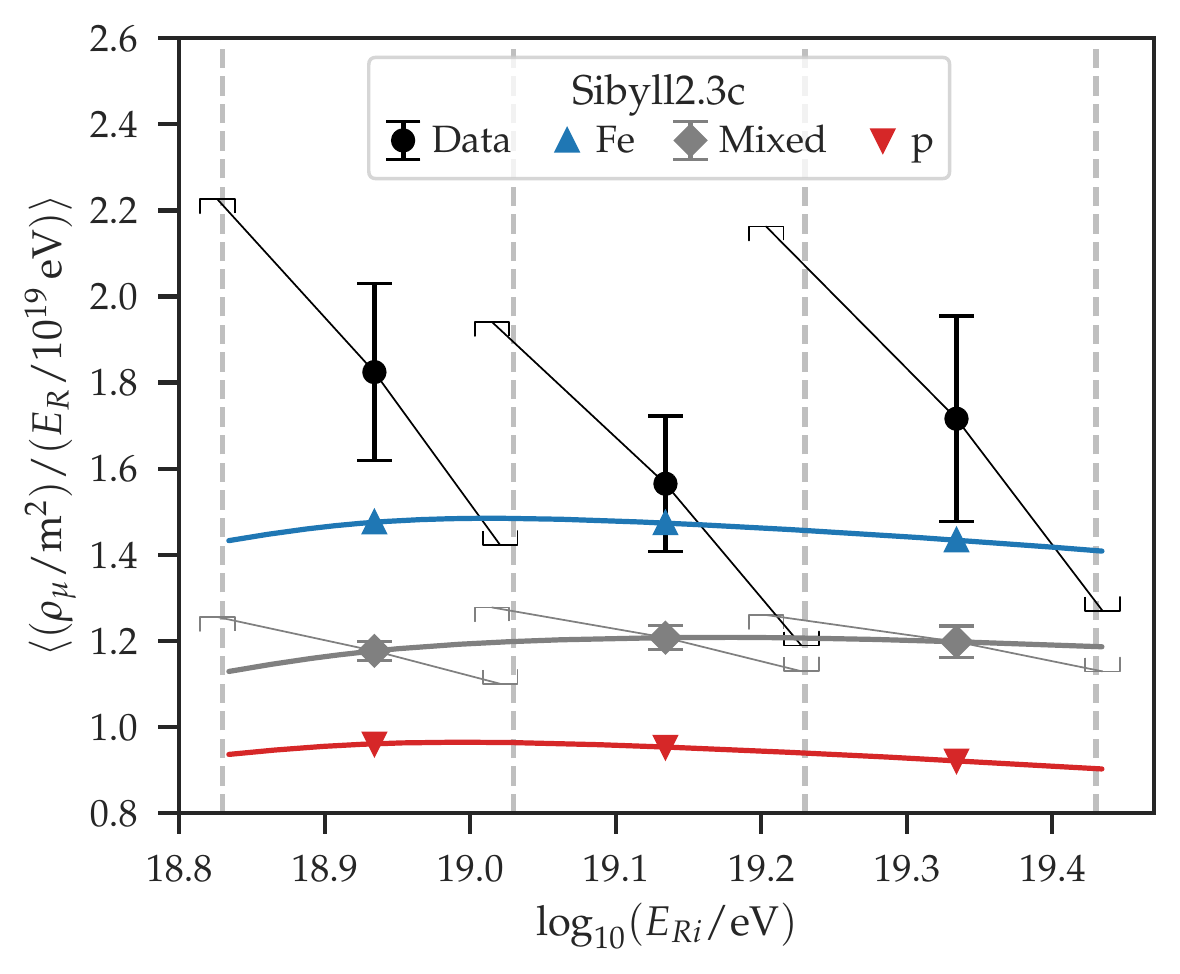}
\includegraphics[width=0.5\linewidth]{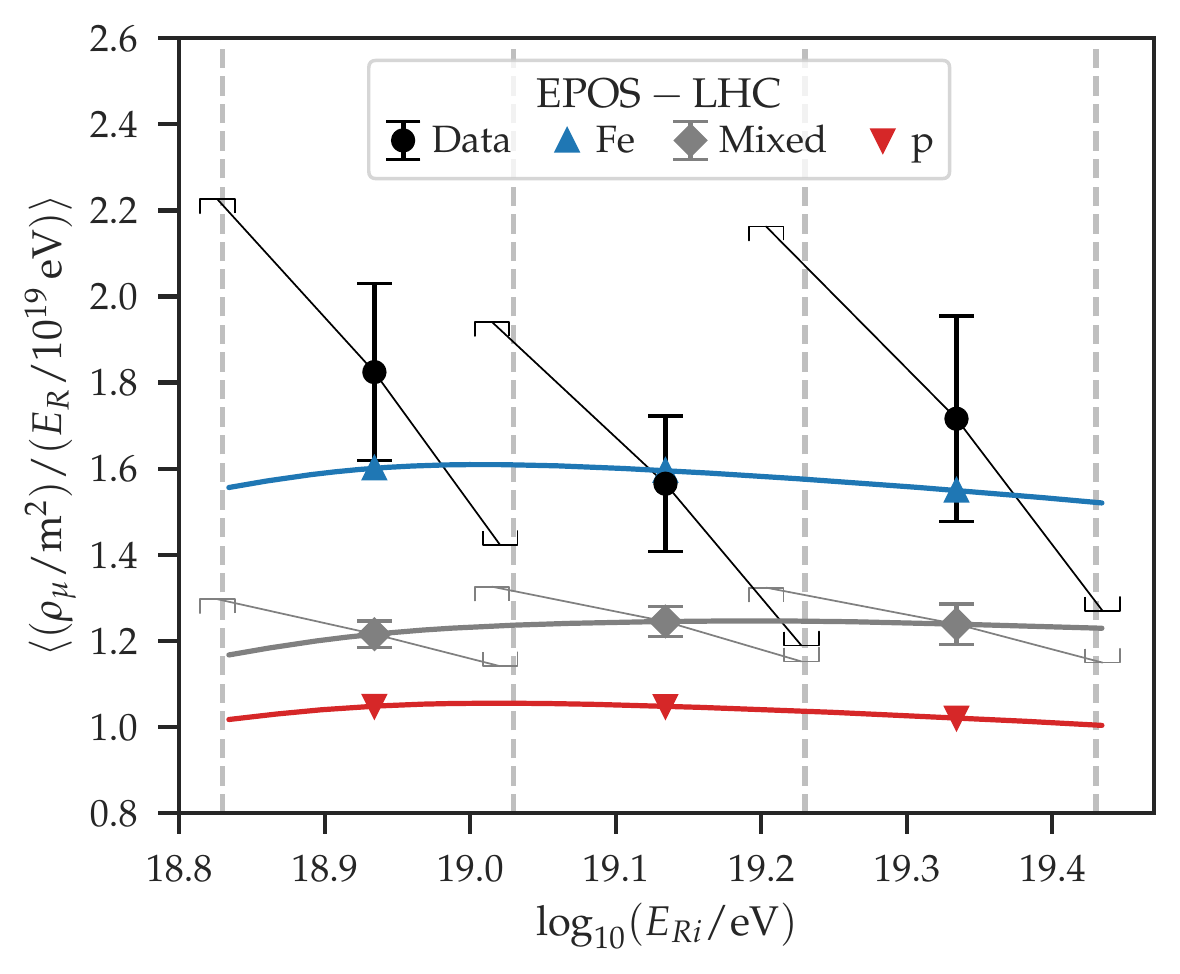}
\begin{center}
\includegraphics[width=0.5\linewidth]{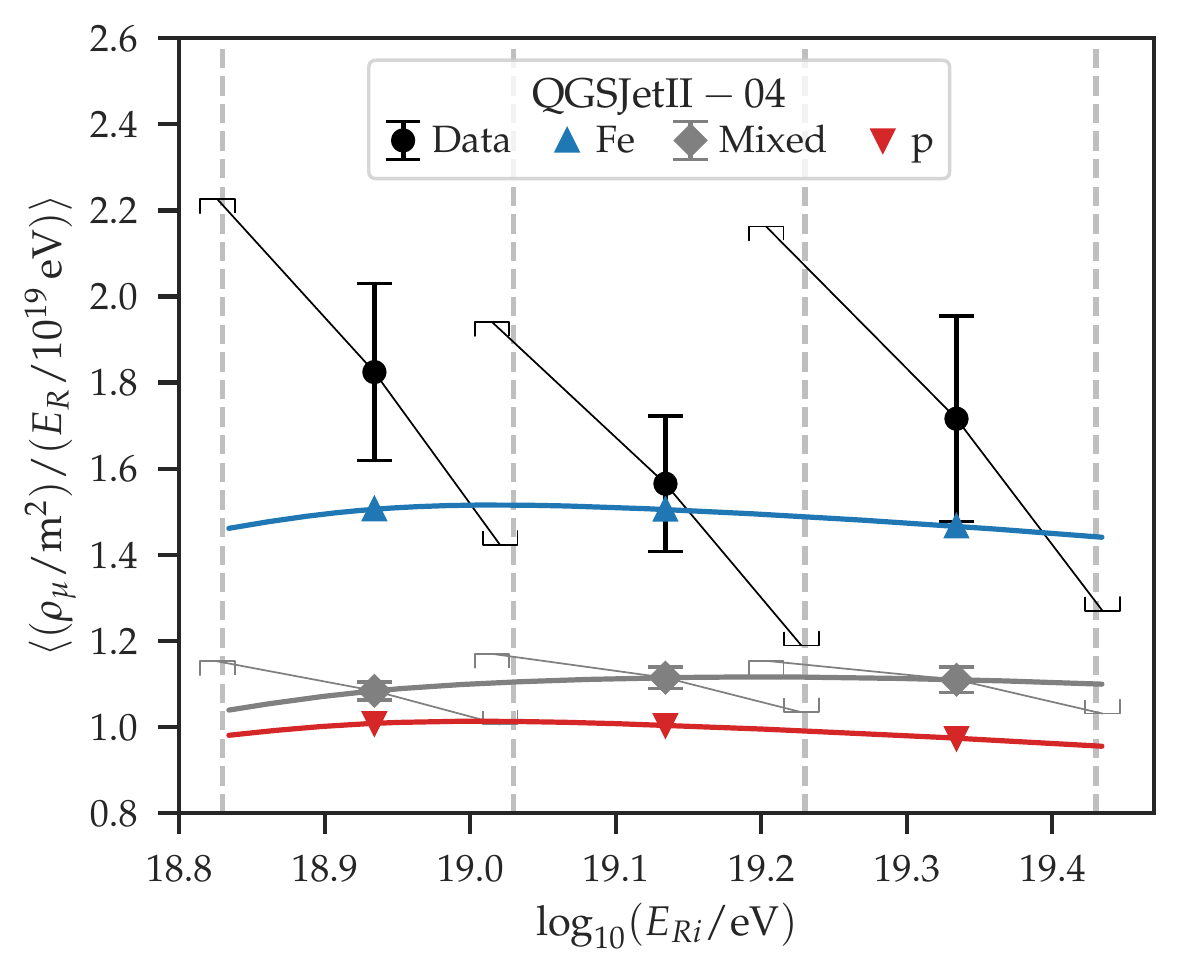}
\end{center}

\caption{Average muon density divided by the reconstructed energy, as a 
function of the logarithm of the reconstructed energy in the center of the 
$i$-th bin. Superimposed to AGASA data points \cite{Shinozaki04} are the 
predictions for proton (red) and iron (blue) primaries, 
and for the mixed composition scenario corresponding to the models Sibyll2.3c 
(top left panel), EPOS-LHC (top right panel), and QGSJetII-04 
(bottom panel). The systematic uncertainties are enclosed by square brackets. 
The vertical dashed lines mark the limits of the reconstructed energy bins. 
\label{fig:RhoMuOverEr}}
\end{figure}
\end{center}

Figure \ref{fig:RhoMuOverEr} shows $\langle\rho_{\mu}/E_R\rangle(E_{Ri})$ 
as a function of the logarithm of the reconstructed energy bin obtained for 
AGASA data, proton and iron simulations, and for the mixed composition 
scenarios. The data from the $120$ events are grouped into three energy bins, each of a binwidth of $\Delta\log_{10}(E/\textrm{eV})=0.2$. The data points constitute the average of $67$, $33$, and $20$ events (from lower to higher energy). The square brackets associated to the AGASA data represent the systematic uncertainties corresponding to the energy (hence they are diagonal) and muon density. The square brackets associated to the mixed composition scenarios include those of the energy, and also include the systematic uncertainties propagated from the mass fractions, which are small in comparison to the first. It can be seen from Fig.~\ref{fig:RhoMuOverEr} that for all models, the AGASA data are compatible with iron nuclei. The AGASA data points lie above the mixed composition scenarios. The lowest energy data point is not compatible with the mixed composition scenarios for all three models. This is also true for the highest-energy data point in the case of QGSJetII-04. However, all the remaining data points are compatible with the mixed composition scnearios when considering total uncertainties. %

\begin{figure}[!ht]
\begin{center}
\includegraphics[width=0.75\linewidth]{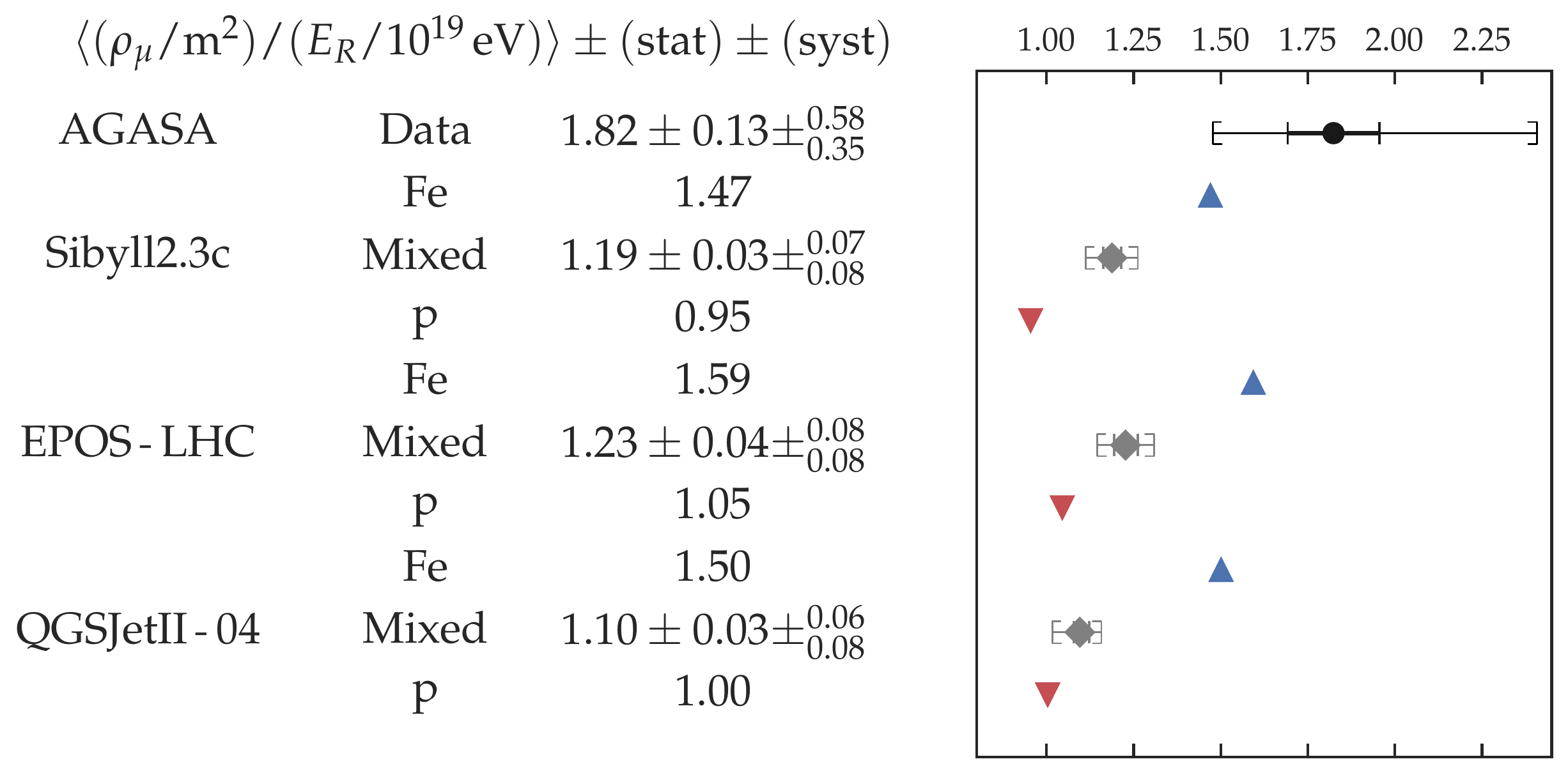}
\caption{Average muon density divided by the reconstructed energy for AGASA 
data and for proton, iron, and mixed composition scenarios, for the three considered models. The obtained values 
are reported in the table (left) and are also plotted (right) on the same line.
The average is taken in the energy range $18.83\,\leq\,\log_{10}(E_{R}/
\textrm{eV})\,\leq\,19.46$. The square brackets show the 
systematic uncertainties.\label{fig:ThreeBin}} 
\end{center}
\end{figure}

To summarize the information in a unique value, we compute $\langle\rho_{\mu} / E_R\rangle$ taking the average in the whole energy range ($18.83\,\leq\,\log_{10}(E_{R}/\textrm{eV})\,\leq\,
19.46$). It is reasonable to compute such average since $\rho_{\mu} / E_R$ is nearly constant within the analyzed energy range. Fig.~\ref{fig:ThreeBin} shows the values of $\langle\rho_{\mu} / E_R\rangle$ computed from AGASA measurements, and the ones corresponding to proton, iron, and mixed composition scenarios for the three considered models.

Fig.~\ref{fig:ThreeBin} shows that the composition inferred from $\langle\rho_{\mu} / E_R\rangle$ 
obtained from AGASA data is compatible with heavy primaries, for the three 
models considered. The values from AGASA data are larger than those of the mixed composition scenarios, and are not compatible within total uncertainties. Considering the upper uncertainties for the mixed composition models, and the lower uncertianties for the AGASA data point, the discrepancies are of $1.9\sigma$ for QGSJetII-04, $1.6\sigma$ for EPOS-LHC, and $1.7\sigma$ for Sibyll2.3c. It follows that the AGASA data support a muon deficit in simulations.

\begin{figure}[!ht]
\begin{center}
\includegraphics[width=0.75\linewidth]{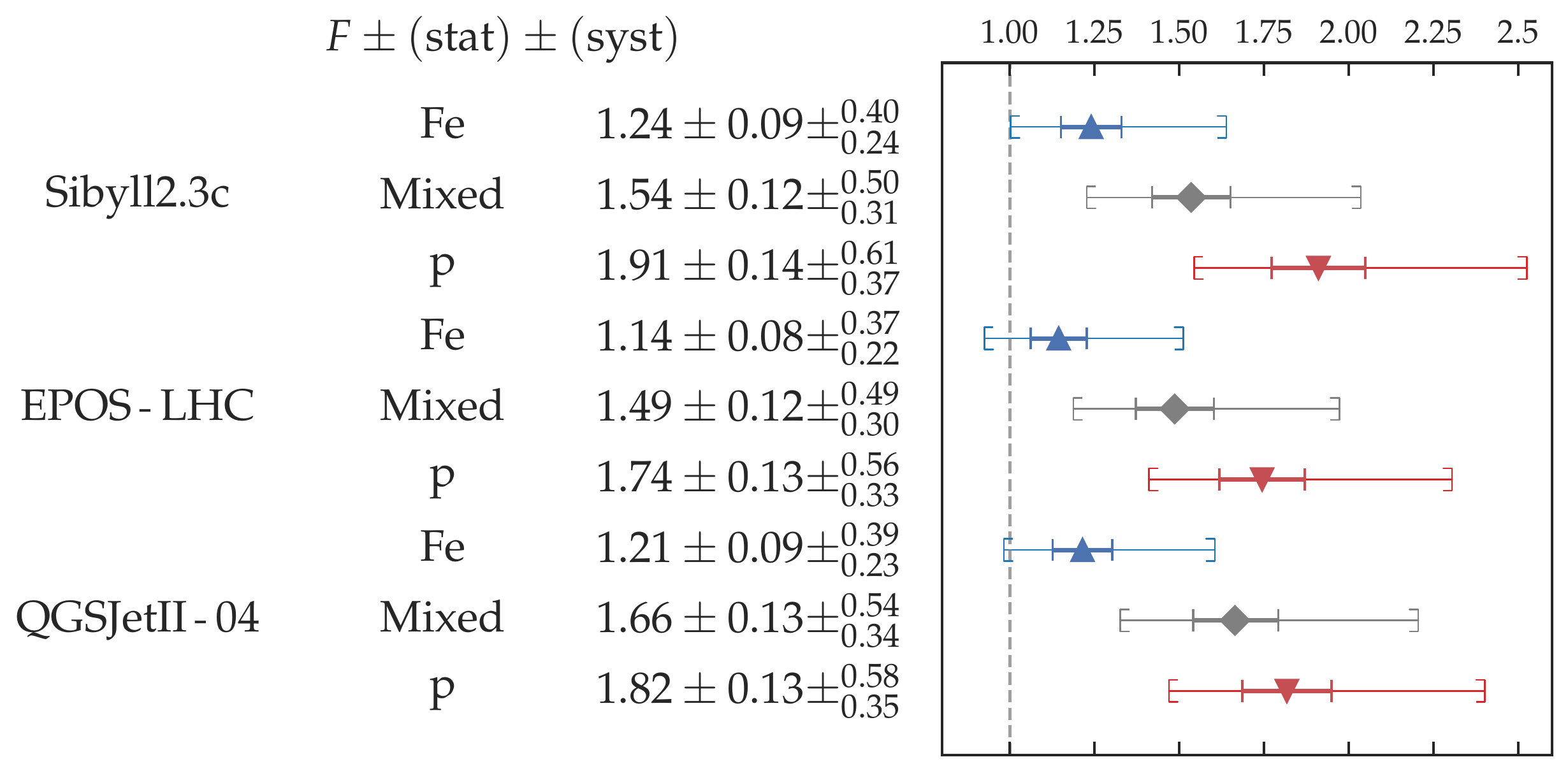}
\caption{Muon density correction factor $F$ corresponding to the single nuclei 
and mixed composition scenarios, for the three considered models. The obtained values are reported in the table 
(left) and are also plotted (right) on the same line. The square brackets correspond to the systematic uncertainties. The energy range under consideration is $18.83\,\leq\,\log_{10}(E_{R}/\textrm{eV})\,\leq\,19.46$.  \label{fig:confint}}
\end{center}
\end{figure}

We therefore quantify the average muon deficit in the reconstructed energy range $18.83\,\leq\,
\log_{10}(E_{R}/\textrm{eV})\,\leq\,19.46$ with a correction factor $F$. This is defined as the ratio between the experimental 
average muon density divided by the energy and the one 
obtained from air shower simulations,
\begin{equation}
\label{eqF}
F = \ddfrac{\langle\rho_{\mu}^{\textrm{data}} / E_R\rangle}{\langle
\rho_{\mu}^{\textrm{S}} / E_R\rangle},
\end{equation}
where S denotes the scenario under analysis, i.e.~S$=$\{mix, p, Fe\}. The 
uncertainties in $F$ are derived by standard error propagation. Fig.~\ref{fig:confint} shows the obtained values of the correction factor $F$, together with their statistic and systematic uncertainties, for the three considered models, and for the single nuclei and mixed composition scenarios. Note that the latter do not overlap with $1$ even considering total uncertainties.

It is interesting to add that the results presented in Figs.~\ref{fig:ThreeBin} and \ref{fig:confint} are essentially independent of the chosen flux parameterization. If the fit to the flux measurements of the Pierre Auger Observatory \cite{Fenu17} are used instead of that of Telescope Array \cite{TAFlux,Gesualdi20}, the values of $\langle\rho_{\mu}^{\textrm{S}} / E_R\rangle$ and $F$ change in less than $\sim 1\,\%$.

The muon deficit found in this analysis is qualitatively compatible with those 
reported by the Pierre Auger \cite{Auger16} and Telescope Array 
Collaborations \cite{TelescopeArray18}. A quantitative comparison to the findings of other experiments can be found in Ref.~\cite{Gesualdi21}.
 
\section{Conclusions}

We analyzed the measurements of the muon density at $1000\,\textrm{m}$ from the shower axis obtained by the AGASA experiment, and compared them to the predictions corresponding to single proton and iron-initiated air shower simulations, as well as four-component mixed composition scenarios. The latter are based on the $X_{\textrm{max}}$ measurements performed by the Pierre Auger Observatory. The simulations were generated with the high-energy hadronic interaction models QGSJetII-04, EPOS-LHC, and Sibyll2.3c. Furthermore, we worked in the reference energy scale defined by the \textit{Spectrum Working Group} \cite{Ivanov17}. To compare data to simulations, we analytically computed the biases introduced by binning in energy and by reconstructing the energy with a broad resolution. In this work, we take into account the systematics in the energy (between $\pm 22\,\%$ to $\pm 26\,\%$) as described in Ref.~\cite{Gesualdi21}.

The AGASA measurements are found to be compatible with iron primaries for all the considered models. The measurements lie above the predictions corresponding to the mixed composition scenarios for the three models. When considering the average throughout the complete energy range, the estimate of $\langle \rho_{\mu}/E \rangle$ from AGASA data does not overlap with those of the mixed composition scenarios. Thus, AGASA data support a muon deficit in simulations. 

We computed a muon density correction factor $F$ in the energy range $18.83\,\leq\,
\log_{10}(E_{R}/\textrm{eV})\,\leq\,19.46$ for the three models considered. For the mixed composition scenarios to be in perfect agreement with AGASA 
measurements, the muon density should be incremented by a factor of $1.49 \pm 
0.11\,\textrm{(stat)} \pm\, ^{0.49}_{0.30}\,\textrm{(syst)}$ for EPOS-LHC, $1.54 \pm 0.12\,
\textrm{(stat)} \pm\, ^{0.50}_{0.31}\,\textrm{(syst)}$ for Sibyll2.3c, and $1.66\pm0.13\,
\textrm{(stat)} \pm ^{0.54}_{0.34}\,\textrm{(syst)}$ for QGSJetII-04. It is worth 
mentioning that the estimated muon deficits are qualitatively in agreement with the ones reported by the Pierre Auger and Telescope Array Collaborations.

\section*{Acknowledgements}
The authors acknowledge the members of the WHISP group, specially K. Shinozaki and H. Dembinski, for useful discussions about the systematic uncertainties of the AGASA data.


\newpage
\begin{thebibliography}{99}
\def\url#1{\href{#1}{#1}}
\def\vyp#1#2#3{\textbf{#1} (#2) #3}
\def\doi#1{\href{http://dx.doi.org/#1}{doi:\,#1}}

\bibitem{Prado19}
R.~R. Prado, 
{\href{http://dx.doi.org/10.1051/epjconf/201920808003}{EPJ Web Conf. \textbf{208}, 08003 (2019)}}.

\bibitem{Auger15}
A. Aab et al. (Pierre Auger Collaboration), 
Phys. Rev. D \textbf{91}, 032003 (2015).

\bibitem{Auger16}
A. Aab et al. (Pierre Auger Collaboration), 
Phys. Rev. Lett. \textbf{117}, 192001 (2016).

\bibitem{TelescopeArray18} 
R. U. Abbasi et al. (Telescope Array Collaboration), 
Phys. Rev. D \textbf{98}, 022002 (2018).

\bibitem{Dembinski19}
H.~P. Dembinski \textit{et al.} for the EAS-MSU, IceCube, KASCADE-Grande, NEVOD-DECOR, Pierre Auger, SUGAR, Telescope Array, and Yakutsk EAS Array collaborations, 
{\href{http://dx.doi.org/10.1051/epjconf/201921002004}{EPJ Web Conf. \textbf{210}, 02004 (2019)}}.

\bibitem{Cazon19}
L. Cazon for the EAS-MSU, IceCube, KASCADE-Grande, NEVOD-DECOR, Pierre Auger, SUGAR, Telescope Array, and Yakutsk EAS Array collaborations, 
PoS (ICRC2019), 214 (2019).

\bibitem{Shinozaki04}
K. Shinozaki and M. Teshima, 
{\href{http://dx.doi.org/10.1016/j.nuclphysbps.2004.10.045}{Nucl. Phys. B (Proc. Suppl.) \textbf{136}, 18 (2004)}}.

\bibitem{Hayashida95}
N. Hayashida \textit{et al.}, 
J. Phys. G: Nucl. Part. Phys. \textbf{21}, 1101 (1995). 

\bibitem{Gesualdi20}
F. Gesualdi, A.~D. Supanitsky, and A. Etchegoyen,
{\href{http://dx.doi.org/10.1103/PhysRevD.101.083025}{Phys. Rev. D \textbf{101}, 083025 (2020)}}.

\bibitem{Ivanov17}
D. Ivanov for the Pierre Auger Collaboration and the Telescope Array Collaboration, 
{\href{https://ui.adsabs.harvard.edu/abs/2017ICRC...35..498I}{PoS (ICRC2017), 498 (2018)}}.

\bibitem{Bellido17}
J. Bellido for the Pierre Auger Collaboration, 
PoS (ICRC2017), 301 (2018).

\bibitem{Dembinski18}
H.~P. Dembinski, 
{\href{http://dx.doi.org/10.1016/j.astropartphys.2018.05.008}{Astroparticle Physics \textbf{102}, 89 (2018)}}.

\bibitem{TAFlux}
C. Jui for the Telescope Array Collaboration, 
{\href{https://ui.adsabs.harvard.edu/abs/2015ICRC...34...35J}{PoS (ICRC2015), 035 (2016)}}.

\bibitem{Takeda03}
M. Takeda \textit{et al.}, 
{\href{http://dx.doi.org/10.1016/S0927-6505(02)00243-8}{Astropart. Phys. \textbf{19}, 447 (2003)}}.

\bibitem{Yoshida95}
S. Yoshida \textit{et al.}, 
{\href{http://dx.doi.org/10.1016/0927-6505(94)00036-3}{Astropart. Phys. \textbf{3}, 105 (1995)}}. 

\bibitem{Gesualdi21}
F. Gesualdi \textit{et al.} for the WHISP Group,
PoS (ICRC2021), 473 (2021).

\bibitem{Fenu17}
F. Fenu for the Pierre Auger Collaboration, 
{\href{https://doi.org/10.22323/1.301.0486}{PoS (ICRC2017), 486 (2018)}}.

\end{thebibliography}
\end{document}